\documentclass[twocolumn,aps,pra,showpacs]{revtex4-1}
\usepackage[latin9]{inputenc}
\usepackage{amsmath}
\usepackage{amssymb}

\begin{document}

\title{Long-lived universal resonant Bose gases}

\author{Shao-Jian Jiang}
\affiliation{Department of Physics and Astronomy, University of British
	Columbia, Vancouver V6T 1Z1, Canada}
\author{Jeff Maki}
\affiliation{Department of Physics and Astronomy, University of British
	Columbia, Vancouver V6T 1Z1, Canada}
\author{Fei Zhou}
\affiliation{Department of Physics and Astronomy, University of British
	Columbia, Vancouver V6T 1Z1, Canada}

\begin{abstract}
Quantum simulations based on near-resonance Bose gases are limited by their short lifetimes due to severe atom losses.
In addition to this, the recently predicted thermodynamical instability adds another constraint on accessing the resonant Bose gases.
In this article, we offer a potential solution by proposing long-lived resonant Bose gases in both two and three dimensions, where the conventional few-body losses are strongly suppressed.
We show that 
the thermodynamical properties as well as the lifetimes of these strongly interacting systems are universal, and independent of short-range physics.
\end{abstract}


\maketitle

Cold atomic gases near resonance offer an ideal platform for the study of strongly interacting quantum systems and have recently attracted considerable interest.
It not only provides a distinctive approach to advancing our understanding of many puzzling quantum many-body problems in various other fields of physics, but also offers intriguing topics of its own.
One outstanding example is the atomic Bose gas near resonance, which is experimentally accessible thanks to the extreme tunability of interatomic interactions.
This allows for experimental studies on Bose gases at large positive scattering lengths~\cite{Papp2008a,Pollack2009b,Navon2011a,Wild2012a,Ha2013a,Rem2013a,Fletcher2013a,Makotyn2014a,fermion}.
Alongside these experimental works, various numerical and theoretical efforts have also been made to analyze this strongly interacting system~\cite{Giorgini1999a,Cowell2002a,Pilati2005a,Song2009a,Diederix2011a,Borzov2012a,Li2012a,Zhou201383,Mashayekhi2013a,Jiang2014a,Yin2013a,Sykes2014a,Kain2014a}.

One of the main constraints that limits the experimental access of resonant Bose gases is their short lifetimes due to few-body losses.
These are recombination processes which lead to the formation of few-body bound states and consequently atom losses.
The minimal number of atoms involved in these processes is three, and the corresponding process is known as three-body recombination.
There have been extensive studies of this process in the few-body context ~\cite{Esry1999a,Kraemer2006a,Knoop2009a,Zaccanti2009a,Pollack2009a,Rem2013a,Fletcher2013a}.
It has been shown that, apart from interference effects, the three-body recombination rate tends to increase rapidly when resonance is approached, which significantly reduces the lifetime of the gas.
This is an intrinsic issue of atomic Bose gases which practically limits the outcome one can get out of this platform near resonance.
Overcoming this difficulty
has become one of the main focuses of recent experimental studies.
For instance, a very interesting attempt applied quenching of the scattering length to access resonance and a local equilibrium was reached~\cite{Makotyn2014a}.

Another complexity of resonant Bose gases is the recently predicted thermodynamical instability near resonance~\cite{Pilati2005a,Borzov2012a,Li2012a,Zhou201383,Jiang2015a}. 
Different from few-body losses, this issue of thermodynamical instability is a many-body effect.
It is characterized by a sign change of the compressibility at a critical scattering length, beyond which the compressibility becomes negative.
In two dimensions (2D), a diagrammatic analysis further implies an anomalous behavior of the compressibility as a precursor of the instability~\cite{Mashayekhi2013a}, which is consistent with a recent experiment on 2D Bose gases~\cite{Ha2013a}.
In three dimensions (3D), however, there is still lack of conclusive experimental evidence for this instability.
Nevertheless, with this predicted many-body instability, accessing resonant gases might run into a sharp phase transition rather than the smooth variation of lifetime predicted by the few-body losses~\cite{Jiang2015a}.

Despite these difficulties, efforts have been made to explore the universal aspects of the gas~\cite{Navon2011a,Wild2012a,Makotyn2014a}. 
Naively, the only relevant length scale of the system at resonance is the interatomic distance, which should universally dictate the low-energy physics of the system.
However,  for resonantly interacting bosons, there are a series of infinite three-body bound states with a discrete scaling symmetry, i.e., the Efimov states~\cite{Efimov1970a}.
The binding energies of the Efimov states further depend on the short-range details of the interatomic potentials.
Therefore, it has been proposed that the many-body properties of the gas would further depend on the non-universal features of the interaction~\cite{Braaten2002,Werner2011,Braaten2011}.
From the few-body point of view, this underlying Efimov physics manifests itself in the three-body recombination process as the interference peaks and dips in the three-body loss rate~\cite{Esry1999a}.
This has been utilized to detect the Efimov effect in experiments~\cite{Kraemer2006a,Knoop2009a,Zaccanti2009a,Pollack2009a}.
On the other hand, in the many-body limit, so far there has been no indisputable experimental evidence of the Efimov effect in resonant Bose gases~\cite{Wild2012a,Makotyn2014a}.
More experimental efforts might be needed to address the issue of universality.

Given the current challenges of studying resonant Bose gases, it is tempting to ask whether there exist long-lived resonant Bose gases  where few-body losses are strongly suppressed and that are also thermodynamically stable and universal.
In this article, we show that such systems can be realized in ultracold Bose gases with anomalous single-particle dispersions, which provide an ideal candidate  for  quantum simulations of the thermodynamics and the universalities of strongly interacting systems.

We first consider a 3D Bose gas with a power-law single-particle dispersion $\epsilon(\boldsymbol{k}) \sim |\boldsymbol{k}|^{3/2+\delta}$ ($\delta$ is a small number) and contact interatomic interactions.
It can be described by the grand-canonical Hamiltonian
\begin{eqnarray}
  \label{eq:1}
H &=&\sum_{\boldsymbol{k}} (\epsilon(\boldsymbol{k}) -\mu) b_{\boldsymbol{k}}^\dagger b_{\boldsymbol{k}}
+ 2 U_0 n_0 \sum_{\boldsymbol{k}} b^\dagger_{\boldsymbol{k}} b_{\boldsymbol{k}}
\nonumber \\
&&+\frac{1}{2} U_0 n_0\sum_{\boldsymbol{k}}
b^\dagger_{\boldsymbol{k}}
b^\dagger_{-\boldsymbol{k}}
+\frac{1}{2}U_0 n_0 \sum_{\boldsymbol{k}} b_{\boldsymbol{k}}b_{-\boldsymbol{k}} \nonumber \\
&&+\frac{U_0}{\sqrt{\Omega}}\sqrt{n_0}
\sum_{\boldsymbol{k'},\boldsymbol{q}} b^\dagger_{\boldsymbol{q}} b_{\boldsymbol{k'}+\frac{\boldsymbol{q}}{2}}
b_{-\boldsymbol{k'}+\frac{\boldsymbol{q}}{2}}+h.c.
\nonumber \\
&&+\frac{U_0}{2\Omega} \sum_{\boldsymbol{k}, \boldsymbol{k'},\boldsymbol{q}} b^\dagger_{\boldsymbol{k}+\frac{\boldsymbol{q}}{2}} b^\dagger_{-\boldsymbol{k}+\frac{\boldsymbol{q}}{2}} b_{\boldsymbol{k'}+\frac{\boldsymbol{q}}{2}}
b_{-\boldsymbol{k'}+\frac{\boldsymbol{q}}{2}}+h.c.
\end{eqnarray}
where $b_{\boldsymbol{k}}^\dagger$ ($b_{\boldsymbol{k}}$) is the bosonic creation (annihilation) operator, with $n_0$ the number density of the condensed atoms, $\mu$ the chemical potential of non-condensed atoms, and $U_0$ the strength of the contact interaction.
$\Omega$ is the volume of the system and the summations are over non-zero momentums.
Note that the momentum summation implicitly contains an ultraviolet cutoff, which is related to the microscopic details of the interatomic interactions.
We set the reduced Planck constant to be unity and choose units so that $\epsilon(\boldsymbol{k}) = k^{3/2+\delta}/2$ where $k = |\boldsymbol{k}|$.
This kind of dispersion can be realized in optical lattices with long-range hopping. 
Here we take the 3D cubic lattice as an example and consider a long-range hopping varying as $t_{ij} \sim |\boldsymbol{r}_i - \boldsymbol{r}_j|^{-9/2-\delta}$, where  $t_{ij}$ denotes the hopping amplitude between lattice sites $i$ and $j$.
After a Fourier transformation, this long-range hopping leads to a dispersion \( \epsilon(\boldsymbol{k}) \sim k^{3/2+\delta} \) near the bottom of the lowest band.
Therefore, the system described by Eq.~\eqref{eq:1} can be realized by loading bosons into such a lattice, with the atom number per lattice site being much less than one.
The tuning of interactions should be achieved by
a single-band resonance driven by the effective mass~\cite{Cui2010a}, different from Feshbach resonances in optical lattices which involve multi-band physics~\cite{Buchler2010a}.

It is worth noticing that long-range hopping can also be potentially realized in other experimental systems.
For instance, in a system of trapped ions~\cite{Friedenauer2008a,Islam2013a}, a power-law long-range spin interaction with tunable power has been produced by coupling the vibrational modes of the ions to their internal degrees of freedom via Raman transitions.
In term of magnons, the bosonic collective excitations of the system, this effectively results in a power-law long-range hopping (see Appendix~\ref{sec:appendix} for more details).
In addition, it has been proposed that in a synthetically spin-orbit coupled Bose gas~\cite{Lin2011a}, a low-energy quartic dispersion can be produced at a critical coupling strength, providing a simulator for the quantum Lifshitz Model~\cite{Po2015a,Wu2015b}.
Therefore, it is very promising that Bose gases with tunable anomalous dispersions can be realized with the aid of optical transitions in cold atom experiments.

One direct consequence of a general power-law dispersion is the breaking of Galilean invariance, which considerably alters the two-body scattering behavior and plays an important role in our analysis.
With the power-law dispersion $\sim k^{3/2+\delta}$, the $T$-matrix for a scattering process with total momentum $\boldsymbol{k}$ and total energy $E$ takes the form
\begin{equation}
\label{eq:4}
T(E,\boldsymbol{k}) = \frac{4 \pi^2 \delta[1 + O(\delta^2)] }{B^{\alpha}-(-E-i0^+)^{\alpha} + f(E,k) \delta},
\end{equation}
where $B$ is the binding energy of the two-body bound state, $\alpha = \frac{3/2 - \delta}{3/2 + \delta}$, and $f$ is a regular function satisfying $f(E,0)=0$.
One can immediately see that $T(E, \boldsymbol{k})$ possesses a completely different form compared with the Galilean invariant case, where $T(E,\boldsymbol{k})$ is simply given by $T(E-k^2/4,0)$.
This is because the relative motion and the center-of-mass motion are inseparable in the absence of Galilean invariance.
Furthermore, the $\boldsymbol{k}$-dependence of $T(E,\boldsymbol{k})$ here is suppressed by an extra power of $\delta$ and is thus irrelevant when $\delta \ll 1$.
It is convenient to define an effective scattering length $\lambda \equiv B^{-\frac{1}{3/2+\delta}}$,
which dictates the low-energy scattering.
It renormalizes the bare parameter $U_0$ via
\begin{equation}
\label{eq:3}
U_0^{-1} + \int \frac{d^3 k}{(2 \pi)^3} \frac{1}{k^{3/2+\delta}} = \frac{1}{4 \pi^2 \lambda^{3/2-\delta} \delta  [1+O(\delta^2)]}.
\end{equation}
The divergent $\lambda$ marks the position of resonance, which is the threshold to form two-body bound states.

To analyze this interacting Bose gas, we apply a self-consistent approach, where the energy density of the system, $E(n_0, \mu)$, for a given $n_0$ and $\mu$ is first calculated.
The chemical potential of a gas with atom density $n$  can then be calculated using the following set of self-consistent equations,
\begin{eqnarray}
\label{eq:6}
\mu_c (n_0, \mu) & = & \frac{\partial E(n_0, \mu)}{\partial n_0},\mu = \mu_c (n_0, \mu) \nonumber\\
n & = & n_0 - \frac{\partial \text{Re} E(n_0,\mu)}{\partial \mu},
\end{eqnarray}
where $\mu_c$ is the chemical potential of the condensed atoms, which should be equal to $\mu$ in equilibrium.
The second line of Eq.~\eqref{eq:6} is the number equation.
The calculation of $E(n_0,\mu)$ is carried out diagrammatically using the effective field theory method~\cite{Borzov2012a, Mashayekhi2013a, Jiang2014a}.
It contains contributions from all $N$-body diagrams with $N=2,3,\cdots$.
In the diagrammatic calculation, instead of $U_0$, the $T$-matrix parametrized by $\lambda$ (Eq.~\eqref{eq:4}) is adopted.

We first discuss the weakly interacting or the dilute limit, which is defined by the condition $\lambda^3 n\delta  \ll 1$.
The dimensionless parameter $\lambda^3 n\delta $ is an analogue of the gas parameter $a^3n$ for 3D Bose gases in free space, where $a$ is the scattering length.
In the dilute limit,
the leading order contribution to 
$\mu$ is given by the Hartree-Fock chemical potential $\mu_{\text{HF}} = 4 \pi^2 \lambda^{3/2-\delta}n\delta $.
The dilute condition states that $\mu_{\text{HF}}$ is much smaller than $B$, which makes it possible to carry out an expansion in the dilute limit.
More specifically, the next-to-leading-order contribution to the chemical potential is proportional to $\mu_{\text{HF}} (\lambda^3 n\delta )^{\alpha}$.
It comes from the $\mu$-dependence of $T$, which results from the energy shift of the non-condensed atoms due to many body effects.
For the quadratic dispersion with $\delta = 1/2$, this correction is proportional to $\mu_{\text{HF}} \sqrt{\lambda^3 n}$, resembling the Lee-Huang-Yang correction for dilute Bose gases~\cite{Lee1957a,Beliaev1958a}.
The next order correction contains an extra factor of $\delta$ but gives the leading order contribution to the imaginary part of the chemical potential, which scales as $\mu_{\text{HF}} (\lambda^3 n\delta )\delta $.
This imaginary part comes from the pole structure in the $T$-matrix and reflects the atom losses due to three-body recombination processes.
Compared with the real part of the chemical potential, the imaginary part is further suppressed by a factor of $\delta$, in addition to the expected suppression by the diluteness parameter.
This indicates an extremely low few-body loss rate and long lifetime.

By solving the self-consistent equations using the method of iteration, we obtain the dilute-limit solution:
\begin{eqnarray}
\label{eq:8}
\mu & = & \mu_{\text{HF}} \left[ 1 + \left( \frac{3}{2} + \frac{8\ln 2 -4}{3} \delta \right) (8 \pi^2 \lambda^3 n\delta )^{\alpha} \right. \nonumber \\
&&  - i 4 \pi (8 \pi^2 \lambda^3 n\delta )\delta  +O( ( \lambda^3 n\delta)\delta^2) \bigg] \nonumber \\
n_0 & = & n \left[ 1 - \frac{1}{2} (8 \pi^2  \lambda^3 n\delta)^{\alpha} + O((\lambda^3 n\delta )\delta ) \right],
\end{eqnarray}
which is asymptotically exact when \( \delta \ll 1 \). 
It only depends on the scattering length and atom density, independent of any ultraviolet parameters.
This suggests the absence of Efimov physics when $\delta$ is small, by contrast to the normally dispersive case with $\delta=1/2$, where Efimov effect is present.

The expansion in Eq.~\eqref{eq:8} breaks down near resonance when $ \lambda^3 n\delta \sim 1$ or $\gg 1$.
However, the presence of the small parameter $\delta$ makes it possible to systematically analyze interacting Bose gases even at resonance.
Following  similar techniques developed in Ref.~\cite{Jiang2014a}, we carry out this analysis  by examining all the diagrams that contribute to the energy density (Readers who are not interested in these technical details can skip to Eq.~\eqref{eq:13}).
Each diagram can be described by two integers, $N$ and $L$, which are  half the number of external condensed atom lines and the number of closed loops respectively.
The corresponding diagram is called an $N$-body $L$-loop diagram.
When $\lambda^3 n\delta  \sim 1$ or $\gg 1$, the energy dependence of the $T$-matrix begins to play an important role.
This makes the self-consistency necessary even at the tree level.
In the tree level, the energy density is $E^{(0)}(n_0,\mu) = n_0^2T(2\mu,0)/2$, where the superscript of $E$ indicates the number of loops. 
Consequently, the self-consistent equation reads,
\begin{equation}
\label{eq:9}
\mu= \frac{4 \pi^2n_0 \delta  [1+O(\delta^2)]}{\lambda^{-(3/2-\delta)} - (-2 \mu - i 0^+)^{\alpha}}.
\end{equation}
In the limit when $ \lambda^3 n\delta \gg 1$, the solution of Eq.~\eqref{eq:9} takes a simple form~\cite{ngtvmu},
\begin{equation}
\label{eq:10}
\mu^{(0)}=(2 \pi^2  n_0\delta)^{1/2+\delta/3}[1+O(\delta)].
\end{equation}
Beyond the tree level, one can easily show that all diagrams are free of ultraviolet divergences after the energy dependence of the $T$-matrix is taken into account.
To analyze the contributions of these diagrams, we take $\mu^{(0)}$ as the leading order result for the chemical potential.
The contribution to the chemical potential from an $N$-body $L$-loop diagram can then be estimated to be
\begin{eqnarray}
\label{eq:12}
\mu^{(L)} & \sim &  
\mu^{(0)}\delta^L.
\end{eqnarray}
which only depends on $L$, as indicated by the superscript of $\mu$ on the left hand side.
One can see that, compared with the tree-level result, contributions from diagrams with $L>0$ are suppressed by extra factors of $\delta$, which validates taking $\mu^{(0)}$ as the leading order result.
Moreover, Eq.~\eqref{eq:12} shows that the contributions from diagrams beyond tree level are organized according to the number of loops, independent of $N$.
This suggests a systematic expansion in terms of $L$ controlled by $\delta$, even though the gas is near resonance or $ \lambda^3 n\delta \gg 1$.

In the limit $ \lambda^3 n\delta \rightarrow \infty$, we solve the self-consistent equations rigorously up to order $O(\delta)$ by summing up all the one-loop diagrams. 
The self-consistent equations
lead to a solution that is universal,
\begin{eqnarray}
\label{eq:13}
\mu & = &  \frac{2 \sqrt{2}}{3} \epsilon_F\delta^{\frac{3/2+\delta}{3}} (1+ 1.22 \delta - i 3.79 \delta + O(\delta^2))  \nonumber \\
n_0 & = & \frac{2}{3} n (1 + 0.40 \delta + O(\delta^2)) 
\end{eqnarray}
where $\epsilon_F = (6 \pi^2 n)^{1/2+\delta/3}/2$ is the Fermi energy defined for the Bose gas with the single-particle dispersion $\epsilon(\boldsymbol{k}) = k^{3/2+\delta}/2$.
It can be seen that the imaginary part of the chemical potential is smaller by a factor of $\delta$ compared with its real part, implying a long-lived Bose gas near resonance.
We shall discuss this in more detail below.

We now carry out a similar analysis for a 2D Bose gas with a nearly-linear dispersion $\epsilon(\boldsymbol{k}) \sim k^{1+\delta}$, which yields similar results.
In the dilute limit, by defining a 2D gas parameter $ \lambda^2 n\delta$, the expansion of $\mu/\mu_{\text{HF}}$ in terms of the gas parameter and $\delta$ has the same form as Eq.~\eqref{eq:8} except for different numerical coefficients.
At resonance, this 2D Bose gas is also shown to be universal and long-lived.
Furthermore, apart from different numerical prefactors, the result expressed in terms of the Fermi energy is organized exactly in the same way as Eq.~\eqref{eq:13}.
Table~\ref{ta:1} summarizes the main results for both cases and the numerical coefficients are shown in Table~\ref{ta:2}.
\begin{table}[t]
\caption{The chemical potentials of Bose gases with dispersion $\epsilon (\boldsymbol{k}) = k^{d/2+\delta}/2$ ($d$ is the spatial dimension and $\delta$ is a small number) in the dilute limit and near resonance in both two ($d=2$) and three dimensions ($d=3$).  Note that they have the same form in both 2D and 3D. The dimensionality-dependent numerical coefficients, $D_1$, $D_2$, $D_3$, $R_1$, $R_2$, and $R_3$, are shown in Table~\ref{ta:2}. Here $\beta = \frac{d/2 - \delta}{d/2 + \delta}$, $\mu_{\text{HF}} \sim \lambda^{d/2-\delta} n \delta$ represents the Hartree-Fock chemical potential, and $\eta = \lambda^d n \delta$ is the dilute gas parameter, with $\lambda$ the effective scattering length and $n$ the atom density. $\epsilon_F \sim n^{1/2+\delta/d}$ is the Fermi energy defined for density $n$.}
\label{ta:1}
\begin{ruledtabular}
\begin{tabular}{lcc}
 & $\text{Re} \mu$ & \(\displaystyle \frac{\text{Im}\mu}{\text{Re}\mu} \) \\
\hline
Dilute limit \\
 ($\eta \ll 1$) & \raisebox{1.5ex}[0pt]{ \( \displaystyle {\mu_{\text{HF}}} [1 + D_1 \eta^{\beta} + D_2 \eta^{\beta} \delta + \cdots]\)} & \raisebox{1.5ex}[0pt]{ \(- D_3 \eta \delta +\cdots \) }\\
Resonance \\ ($\eta \rightarrow \infty $) & \raisebox{1.5ex}[0pt]{  \( \displaystyle {\epsilon_F} [  R_1\delta^{\frac{d/2+\delta}{d}}( 1 + R_2 \delta + \cdots)] \)} & \raisebox{1.5ex}[0pt]{ \( - R_3 \delta + \cdots\)} 
\end{tabular} 
\end{ruledtabular}
\end{table}

\begin{table}[t]
\caption{Numerical coefficients in solutions of the chemical potential in 2D and 3D. The solutions with these coefficients are shown in Table~\ref{ta:1}.}
\label{ta:2}
\begin{ruledtabular}
\begin{tabular}{lcc}
 & 2D & 3D \\
\hline
$D_1 $ & $12\pi$ & $12\pi^2$ \\
$D_2$ & $-8\pi (2+5\ln 2 + 3 \ln \pi)$ & \(\displaystyle -16\pi^2 \left(\frac{2}{3} + \frac{5 \ln 2}{3} + 2 \ln \pi \right) \)\\
$D_3$ & $48 \pi^2$ & $32 \pi^3 $\\
\hline
$R_1$ & \( \displaystyle \frac{2 \sqrt{3}}{3} \) & \(\displaystyle \frac{2 \sqrt{2}}{3} \) \\
$R_2$ & $2.08$ & $1.22$\\
$R_3$ & $5.89$ & $3.79$
\end{tabular}
\end{ruledtabular}
\end{table}
It can be seen that the equations of state at resonance have the same universal form in both 2D and 3D.
Since resonance corresponds to a fixed point of the running two-body coupling constant, this implies that the 3D Bose gas with dispersion $\epsilon(\boldsymbol{k}) \sim k^{3/2+\delta}$ and the 2D Bose gas with dispersion $\epsilon(\boldsymbol{k}) \sim k^{1+\delta}$ belong to the same universality class near resonance.
This can be further generalized to a series of Bose gases with $\epsilon(\boldsymbol{k}) \sim k^{d/2+\delta}$, where $d$ is the dimensionality.
A detailed discussion of this universality class in different spatial dimensions will be presented in an upcoming article.

One important feature of the systems discussed above  is their long lifetimes.
The lifetime of the gas is deduced from the imaginary part of the chemical potential, which can be generally applied both in the dilute limit and near resonance.
In the dilute limit, this can be understood by considering the condensate wave function $\Psi e^{-i \mu t}$, whose time evolution is given by $e^{-i\mu t}$.
A negative imaginary part of the chemical potential implies  a damping effect and the lifetime of the condensate is given by $|\text{Im}\mu|^{-1}$.
The leading order contribution to $\text{Im}\mu$ comes from the three-body scattering processes and can be directly connected to the three-body loss rate. 
The characterization of the lifetime using $\text{Im}\mu$ then reduces to the standard description in terms of the few-body losses.
In contrast, near resonance, contributions from all $N$-body processes with $N=3,4,5, \cdots$ are expected to be parametrically inseparable.
As a result, the relation between the lifetime and the three-body loss rate cannot be established straightforwardly.
In this case, the imaginary part of the chemical potential is due to collective effects and naturally encodes the information of the lifetime.
Its real part, on the other hand, serves as the characteristic many-body energy scale.
Therefore, a long-lived Bose gas can be defined by $|\text{Im} (\mu)/ \text{Re}(\mu)| \ll 1$, which guarantees the realization of equilibrium before significant losses occur.
According to Eq.~\eqref{eq:13} and the third column of Table.~\ref{ta:1}, $|\text{Im} (\mu)/ \text{Re}(\mu)| \sim \delta $, which is parametrically small.
This not only indicates that those gases have long lifetimes at resonance,
but also justifies our method since equilibrium is necessary for the self-consistent approach.

Apart from the short lifetime, the thermodynamical instability that has been pointed out for 2D and 3D Bose gases with normal dispersions~\cite{Borzov2012a, Mashayekhi2013a, Jiang2015a}  is not present in the systems we proposed here.
This can be seen by examining the compressibility, which can be calculated via $\kappa = (\partial \text{Re}\mu / \partial n)^{-1}$.
Table~\ref{ta:1} and Table~\ref{ta:2} show that $\text{Re}\mu \sim \sqrt{n}$ and $\text{Re}\mu>0$, which leads to a positive compressibility near resonance, indicating the thermodynamical stability of the gas.

These long-lived and stable resonant Bose gases are universal, independent of microscopic details of atomic interactions.
This is shown by the equation of state ($\text{Re} \mu(n)$) which only depends on the Fermi energy, an energy scale set by the interatomic distance.
Moreover, the dynamics, i.e., the lifetime ($|\text{Im} \mu|^{-1} \sim \epsilon_{F}^{-1}$) is also universal, again only depending on the Fermi energy or the density.  This aspect is coincidentally consistent with the experimental result in Ref.~\cite{Makotyn2014a}, where the timescale of the dynamics of a 3D Bose gas at resonance had been scaled in terms of the Fermi time, $t_F = \epsilon_F^{-1}$.
Therefore our results might shed further light on future studies of the dynamics of Bose gases.

The finite lifetime of the gases of scattering atoms also indicates that there are other cluster states in the lower branch.
However, the nature of those states remains unknown, and it would be interesting to investigate the connection between those states and the known cluster states of quadratically dispersive bosons~\cite{Platter2004a,Stecher2009a,Stecher2010a}.

On the other hand, the systems are strongly interacting near resonance.
This can be seen from the depletion fraction of the atoms which takes a universal value of $1/3$ and is not parametrically small. 
Additionally, the scaling behavior of the chemical potential near resonance ($\mu \propto \sqrt{\delta}$) is distinctly different from the linear dependence ($\mu \propto \delta$) in the dilute limit where the systems are weakly interacting.

In summary, we demonstrate that long-lived resonant Bose gases can be potentially realized in optical lattices with long-range hopping in both two and three dimensions.
We show this by proposing Bose gases that have parametrically long lifetimes and are thermodynamically stable at resonance.
We further prove that they are universal and strongly interacting, which makes them promising candidates for simulating universal physics of strongly interacting many-body systems.

\begin{acknowledgments}
This work is in part supported by the Canadian Institute for
Advanced Research and the Natural Sciences and Engineering Research Council of Canada (Contract No. 288179).
\end{acknowledgments}

\appendix*
\section{Effective long-range hopping of  magnons in trapped ion systems}
\label{sec:appendix}

In the trapped ion systems, power-law long-range spin interactions have been realized by coupling the vibrational modes of the ions to their internal degrees of freedom via Raman transitions~\cite{Friedenauer2008a,Islam2013a,Senko2015a}. 
Here, we take the following spin-$S$ long-range XXZ Hamiltonian as an example,
\begin{eqnarray}
\label{eq:2}
H & = & \sum_{i<j} J_{ij} \left( \sigma_i^x \sigma_j^x + \sigma_i^y \sigma_j^y + \lambda \sigma_i^z \sigma_j^z \right) \nonumber \\ 
&& + h \sum_i \sigma_i^z,
\end{eqnarray}
where $\sigma^x,\sigma^y, \sigma^z$ are Pauli matrices and $i,j$ denote lattice sites.
$J_{ij} = J_0/|i-j|^{\gamma}$ represents the long-range spin coupling, where $J_0$ is the coupling constant, $|i-j|$ is the distance between lattice sites $i$ and $j$, and $\gamma$ is a tunable number.

Under the standard Holstein-Primakoff transformation~\cite{Mahan2000a,Zhou2004a}, the Hamiltonian [Eq.~\eqref{eq:2}] can be rewritten in the magnon representation as
\begin{eqnarray}
\label{eq:5}
H & = & 4S \sum_{i \neq j} J_{ij} (a_i^{\dagger} a_j  -  \lambda a_i^{\dagger} a_i) - 2 h \sum_i a_i^{\dagger} a_i   \nonumber\\ 
 &  & -  \sum_{i \neq j} J_{ij} (a_i^{\dagger} a_j^{\dagger} a_i a_i + h.c. - 2 \lambda a_i^{\dagger} a_j^{\dagger} a_i a_j) \nonumber\\
 &  & + O(S^{-1}),
\end{eqnarray}
where $a_i^{\dagger} (a_i)$ is the creation (annihilation) operator of magnons on lattice site $i$.
One can see that the long-range spin interactions induce an effective long-range hopping of magnons with the same power-law behavior.

\end{document}